\documentclass[10pt, twocolumn, prd,amsmath,amssymb,floatfix, notitlepage, longbibliography]{revtex4-1}
\usepackage{mathtools}
\usepackage[dvips]{graphics}
\usepackage{bm}
\usepackage{epsfig}
\usepackage{enumerate}
\usepackage{subfigure}
\usepackage{color}
\usepackage{graphicx}
\usepackage[dvipsnames]{xcolor}
\usepackage{graphicx}
\usepackage[colorlinks,citecolor=blue,linkcolor=blue,urlcolor=blue,plainpages=false,pdfpagelabels]{hyperref}

\begin{document} 
% ==============================================================================

\title{Phase stiffness in flat-band superconductors with nodal pairing}

\author{
A.~A.~ Zyuzin and 
A.~Yu.~Zyuzin\textsuperscript{1}
\\
\textsuperscript{1}A.~F.~Ioffe Physical - Technical Institute, 194021 St Petersburg, Russia
}

% ==============================================================================

\begin{abstract}
We study Bogoliubov quasiparticle spectrum in a two-band system with momentum-dependent hybridization between a dispersive band and a flat band. The interplay between the interband mixing and intraband Cooper pairing may give rise to a parabolic node in the spectrum of flat band quasiparticles, resulting in a quadratic temperature dependence of the superconducting phase stiffness at low temperatures.
We also comment that nonmagnetic disorder induces Machida-Shibata deep subgap resonances suggesting the sensitivity of flat-band superconductivity to disorder.
\end{abstract}
\maketitle
%%%%%%%%%%%%%%%%%
\section{Introduction}

The Cooper pairing may be enhanced due to increased electronic density of states when the chemical potential is tuned close to band-structure singularities such as extended van Hove points \cite{Dzyaloshinskii_review, Volovik_1994, Yudin_PhysRevLett} or bands with strongly suppressed quasiparticle kinetic energy (flat-bands) \cite{Volkov_heavy_fermion, Khodel_Shaginyan, Nozieres, PhysRevLett_Masanori, Physica_C_Furukawa, PhysRevB_Kopnin_Heikkila_Volovik}, for reviews see \cite{Volovik_review, balents_review}. 

However, besides enhanced Cooper pairing temperature, superconductivity requires phase coherence which is established due to kinetics of quasiparticle correlations being suppressed in the flat-band regime. Thus, one of the questions arising from these studies is how superconductivity evolves when Cooper pair formation and global phase coherence become strongly distinct processes and appear at largerly different temperatures, for a review \cite{Nature_Sacepe_Feigelman}. This situation may resemble, for example, the crossover region in between weak and strong coupling regimes of superconductivity \cite{JLTP_Schmitt, Nozieres, BCS_BEC1, BCS_BEC2} or disordered superconductors close to localization \cite{Ma_Lee_PhysRevB, Randeria_PhysRevB, Feigelman}. In both cases the superconducting state becomes sensitive to phase disorder and competing states.

However, flat bands often coexist with dispersive bands in real materials, and the correlation between these subsystems may results to a finite supercurrent, \cite{Kopnin_MAIN, Peotta_Torma}, helping to stabilize phase coherence of Cooper pairs within the flat band.
Such correlations may also arise due to intraction-driven mechanism analogous to the phase coherence of excitons in quantum Hall bilayers, where the kinetic energy of particles is quenched by a strong magnetic field \cite{Yang_excitons_FB}.

The idea of Cooper pairing in quasi-flat bands have recently attracted attention, particularly motivated by the observation of superconducting phases in graphene-based heterostructures  \cite{Exp_TBG_superconductor, Dean_science, Yazdani_nature}. For example, experiments have reported a low-temperature power-law dependence of the superconducting phase stiffness in twisted bilayer \cite{Tanaka_2025} and trilayer graphene \cite{Banerjee_2025}. Given that these materials are multiband, it is natural to ask about the role of possible multigap pairing on the quasiparticle spectrum and on the temperature dependence of the phase stiffness.

Here, we consider a minimal two-band model composed of the dispersive parabolic band hybridized with the flat band \cite{Khalf_model_Hamiltonian, Vishwanath_model_Hamiltonian}, representing a limiting case of more general Volkov-Pankratov model \cite{VP_model, BHZ_model}. The fine-tuning of momentum-dependent hybridization between the bands supports formation of the perfectly flat band gapped from the dispersive band. This model might be relevant for heavy-fermion-type quantum systems, as discussed in \cite{Song_PhysRevLett, Khalf_model_Hamiltonian}. 

We find that, due to the interplay between the pairing interaction and the momentum-dependent hybridization, the quasiparticle spectrum develops a parabolic nodal structure, leading to a quadratic temperature dependence of the superconducting phase stiffness in the flat band. We also comment that nonmagnetic impurities can induce deep in-gap Machida - Shibata \cite{Machida_Shibata, Balatsky_RevModPhys} bound states, noting that Cooper pairing in flat-band systems may be sensitive to the presence of disorder.

\section{Model}
\begin{figure}[t!]
    \centering
    \includegraphics[width=7cm]{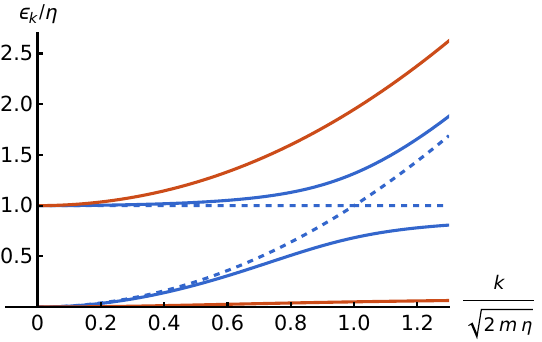}
    \caption{Plot of the band dispersion $\epsilon_{{\bm k}}$ in Eq. (\ref{Normal_disp}), normalized by $\eta$ for different values of the parameter $2 m v^2/\eta = 0, 0.1, 0.9$. As this parameter increases, the lower band flattens.}
    \label{fig:myplot}
\end{figure}

To model the low-energy electronic structure of a two-dimensional system hosting flat band, we consider a two-band Hamiltonian with momentum-dependent interband hybridization of the form, \cite{VP_model, BHZ_model, Khalf_model_Hamiltonian, Vishwanath_model_Hamiltonian, Song_PhysRevLett}:
\begin{eqnarray}
H_s({\bm k})=  \left(\begin{array}{cc}
\frac{{\bm k}^2}{2m}& v(s k_x - i k_y)  \\
v(s k_x + i k_y) & \eta \\
\end{array}\right),
\end{eqnarray}
where $m$ is the effective mass of the dispersive band, $v$ characterises the interband coupling, $\eta>0$ defines the band gap energy, and $s=\pm$ labels two spin projections. The natural units $\hbar= k_{\rm B}=1$ are used throughout the paper. 
Diagonalising this toy-model Hamiltonian yields the band dispersion (the spin degeneracy is implied)
\begin{equation}\label{Normal_disp}
\epsilon_{{\bm k},\pm} = \frac{k^2}{4m} + \frac{\eta}{2} \pm \sqrt{\left(\frac{k^2}{4m} + \frac{\eta}{2}\right)^2 + \left(2mv^2 - \eta\right)\frac{k^2}{2m}},
\end{equation} 
which is also shown in Fig. (\ref{fig:myplot}). Focusing on the lower band, one finds
$
\epsilon_{{\bm k},-} \simeq \left(\eta-2 m v^2 \right) k^2/(2m \eta+ k^2)
$.
This expression reveals that when the parameters are fine-tuned $\eta = 2 m v^2$, the lower band becomes exactly flat 
pinned at zero energy. The large-momentum cutoff $k_0$ must be introduced to regularize such flat-band states.

The question on Cooper pairing in the similar quasi-flat band regime when, $\eta \neq 2 m v^2$, although within momentum-independent interband hybridization model, was investigated in Ref. \cite{Zyuzin_quasi_flat} highlighting the usual formation of two-particle bound states, preformed Cooper pair crossover temperature in two-dimensions, and fully gapped quasiparticle excitations. 

In contrast, here we consider momentum dependent interband hybridization. This coupling allows us to consider a fine-tuned limiting case $\eta = 2 m v^2$, in which the system consists of a perfectly flat band, $\epsilon_{{\bm k}, -} = 0$, separated by a gap $\eta$ from a parabolic conduction band $\epsilon_{{\bm k}, +} = \eta + k^2/2m$. This limit is particularly interesting because it maximises the effect of the density of states on the Cooper pairing crossover temperature, while the interband hybridization still contributes to the phase stiffness. 
However, we note that this is a fine-tuned point in parameter space and may be fragile to perturbations as it is not symmetry protected.

Rescaling the Hamiltonian by $\eta$, we obtain: 
\begin{eqnarray}\label{normalized_Hamiltonian}
H_s({\bm k})=  \eta \left(\begin{array}{cc}
\lambda^2 {\bm k}^2 & \lambda (s k_x - i k_y)  \\
\lambda (s k_x + i k_y) & 1 \\
\end{array}\right),
\end{eqnarray}
where $\lambda = v/\eta=1/2mv$.
It is convenient to write the Green function of electrons through two-band separate contribution:
\begin{equation}\label{Green_function}
G_{s}({\bm k},\omega_n) = \frac{1- \frac{\boldsymbol{\sigma} {\bm h}_s({\bm k})}{h_s({\bm k})}}{2i\omega_n} + 
\frac{1}{2}\frac{1+ \frac{\boldsymbol{\sigma}{\bm h}_s({\bm k})}{h_s({\bm k})}}{i\omega_n - \eta (1+ \lambda^2 k^2)},
\end{equation}
where $\omega_n = \pi T(2n+1)$ is the Matsubara frequency at temperature $T$ and $n\in \mathbb{Z}$, $\boldsymbol{\sigma} = (\sigma_1, \sigma_2, \sigma_3)$ are the Pauli matrices acting in the spin space, and ${\bm h}_s({\bm k}) = \eta [\lambda s k_x, \lambda k_y, (\lambda^2k^2-1)/2]$ with $h_s({\bm k}) \equiv |{\bm h}_s({\bm k}) |$. It can be seen that although the kinetic contribution to the spectrum of the flat band electrons vanishes, noting the denominator in first term in (\ref{Green_function}), there is yet a nontrivial momentum dependence of the Green function matrix structure. This results in the non-vanishing spatial dependence of the flat-band quasiparticle Green function. For example, this is similar to the case of three-band semimetal case in which a flat-band intersects a dispersive band, as discussed in \cite{Zyuzin_Zyuzin}.

%%%%%%%%%%%%%%%%%%
%%%%%%%%%%%%%%%%

\section{Cooper pairing in flat band}

\subsection{Quasiparticle spectrum}
Let us now consider local s-wave-symmetric Cooper pairing in the spin-singlet, band-triplet channel. 
The two-particle bound state can be found from the solution of the eigenvalue equation 
\cite{Sablikov_PhysRevB, Zyuzin_quasi_flat}:
\begin{align}\nonumber\label{2_particle_ham}
&H_{+, \alpha\nu}(\hat{{\bm k}}_1) \Psi_{\nu\beta}({\bm r}_1, {\bm r}_2) + H_{-, \beta \nu}(\hat{{\bm k}}_2) \Psi_{\alpha\nu}({\bm r}_1,{\bm r}_2) \\
&= [E \delta_{\alpha\beta}- U_{\alpha\nu}({\bm r}_1, {\bm r}_2)] \Psi_{\nu\beta}({\bm r}_1,{\bm r}_2),
\end{align}
where $\hat{{\bm k}}_{1,2} = -i\boldsymbol{\nabla}_{{\bm r}_{1,2}}$ is a momentum operator; $\Psi_{\nu\beta}({\bm r}_1,{\bm r}_2)$ is the two-particle wave-function being a $2\times 2$ matrix; $\alpha,\beta,\nu$ are the band indices; and 
$U({\bm r}_1, {\bm r}_2) = - {\rm diag}[V_1,V_2] \delta ({\bm r}_1-{\bm r}_2)$ is the local attraction potential, with positive $V_{1,2}$ differ between the bands. This model implies that the pairing potential spatial scale is smaller than $\lambda$ and $k_0^{-1}$.

Equation (\ref{2_particle_ham}) can be rewritten in a more transparent form as
$H({\bm r}_1, {\bm r}_2) \Phi({\bm r}_1, {\bm r}_2) = [E- \tilde{U}({\bm r}_1, {\bm r}_2)] \Phi ({\bm r}_1, {\bm r}_2)$, where now 
$ \Phi = ( \Psi_{11}, \Psi_{12},\Psi_{21}, \Psi_{22})^{\rm T}$,
$\tilde{U}({\bm r}_1, {\bm r}_2) = - {\rm diag}[2V_1,V_1+V_2,V_1+V_2, 2V_2] \delta ({\bm r}_1-{\bm r}_2)$, and
\begin{eqnarray}
H= \eta \left(\begin{array}{cccc}
\lambda^2 (\hat{{\bm k}}^2_1+\hat{{\bm k}}^2_2) & -\lambda \hat{k}_{2,+}& \lambda \hat{k}_{1,-}&0 \\
-\lambda \hat{k}_{2,-}&1+ \lambda^2 \hat{{\bm k}}^2_1 &0 & \lambda \hat{k}_{1,-} \\
\lambda \hat{k}_{1,+}&0&1+ \lambda^2 \hat{{\bm k}}^2_2 & -\lambda \hat{k}_{2,+} \\
0&\lambda \hat{k}_{1,+} & -\lambda \hat{k}_{2,-}&2 \\
\end{array}\right),~~~~
\end{eqnarray}
with $\hat{k}_{i,\pm} =\hat{k}_{i,x} \pm i\hat{k}_{i,y} $. We focus on bound states with zero center-of-mass momentum and consider a wide flat band case $\lambda k_0>1$ within a "weak-potential limit", in which ${\rm max}(V_1, V_2)/(\lambda^2 \eta) \ll 1$. 

In the absence of the potential, the three eigenvalues are $0$, $\eta(1+\lambda^2k^2)$, and $2\eta(1+\lambda^2k^2)$ corresponding, respectively, to two particles both from the flat band, from different bands, and both from the dispersive band. 
In what follows we shall focus on the lowest-energy bound states, $E = - \frac{k_0^2}{2\pi} \max (V_2,  V_1/\lambda^2 k_0^2 )$,
which is a well known result for the two-particle bound state energy in flat-band being proportional to the interaction constant, obtained in the context of strong coupling superconductors \cite{Khodel_Shaginyan} and superconductivity in quantum Hall states \cite{MacDonald_quantumhall}.

To proceed, following Ref. \cite{Bashkin_tc, PhysRevLett_Randeria}, within mean field approximation ansatz, one introduces the intraband pairing gap functions $\Delta_{1,2}$, constructed on the bare quasiparticle operators of the lower (1) and upper (2) bands. The BdG Hamiltonian takes the form
\begin{eqnarray}
H_{\rm BdG}({\bm k}) =  \left(\begin{array}{cc}
H_{+}({\bm k})-\mu  & \Delta  \\
\Delta^{*} & -H^*_{-}(-{\bm k})+\mu \\
\end{array}\right),
\end{eqnarray}
with
\begin{eqnarray}
\Delta =  \left(\begin{array}{cc}
\Delta_1  & 0  \\
0 & \Delta_2 \\
\end{array}\right).
\end{eqnarray}
The pairing parameters $|\Delta_{1,2}|$ are proportional to the amplitudes of the binding energies $V_1/\lambda^2$ and $V_2 k_0^2$, respectively.
For example, at charge neutrality (which might correspond to the strong-coupling regime of superconductivity when the pairing amplitude is comparable to or exceeding the Fermi energy), $\mu=0$, in the case $V_1=0$, one obtains $|\Delta_1|=0$ and using
\begin{equation}
 \Delta_2 = V_2 T\sum_n \int \frac{d^2k}{(2\pi)^2} \frac{\lambda^4 k^4}{(1+\lambda^2k^2)^2} \frac{\Delta_2}{\omega_n^2 + \frac{\lambda^4k^4|\Delta_2|^2}{(1+\lambda^2k^2)^2}}
\end{equation}
we get
$|\Delta_2| = \frac{V_{2} k_0^2}{8\pi} \tanh(|\Delta_2|/2T) + O(\ln|\lambda k_0|/\lambda k_0)$, with $\lambda k_0\gg 1$.

\begin{figure}[t!]
    \centering
    \includegraphics[width=7cm]{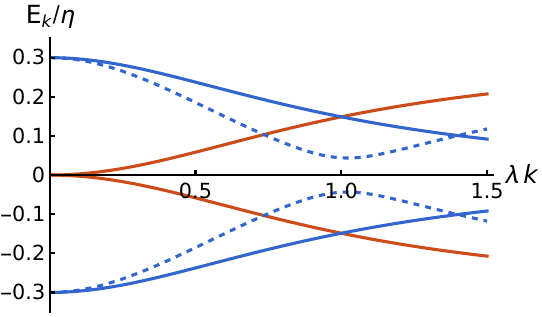}
    \caption{Plot of the lower band dispersion in Eq. (\ref{Dispersion_SC}), $E_{{\bm k},-}$, normalized by $\eta$ as a function of $\lambda k$ for different values of the parameters: dashed curves correspond to the $\pi$-shift case $\Delta_1 = -\Delta_2=0.3 \eta$; the curves that vanish at the origin correspond to $\Delta_1=0, \Delta_2=0.3 \eta$; and the remaining ones to $\Delta_1=0.3 \eta, \Delta_2=0$ vanishing at large momenta as $k^{-2}$.}
    \label{fig:myplot2}
\end{figure}

Let us futher analize the spectrum of quasiparticles in the presence of the gap-functions, $\Delta_1$ and $\Delta_2$. At charge neutrality the quasiparticles spectrum reads
\begin{eqnarray}\nonumber\label{Dispersion_SC}
2 E_{{\bm k},\pm}^2 &=& |\Delta_1|^2+|\Delta_2|^2+\eta^2(1+ \lambda^2 k^2)^2 \pm
\\\nonumber
&\pm& \{[|\Delta_1|^2+|\Delta_2|^2+\eta^2(1+ \lambda^2 k^2)^2]^2
\\
&-& 4 |\Delta_1|^2|\Delta_2|^2 -4 \eta^2 |\Delta_1+\lambda^2 k^2 \Delta_2|^2 \}^{1/2}.~~~
\end{eqnarray}
This dispersion is shown in Fig. (\ref{fig:myplot2}).
For a symmetric case, $\Delta_1 \equiv \Delta_2$, the quasiparticle spectrum simplifies. Then we get an expression for an upper $E_{{\bm k},+} = \pm\sqrt{|\Delta_2|^2 + \eta^2(1+\lambda^2k^2)^2}$ and lower $E_{{\bm k},-}= \pm |\Delta_2|$ bands (where $\pm$ denotes particle-hole symmetric branches).

However, asymmetry in $|\Delta_{1,2}|$ leads to the momentum dependence of the spectrum of quasiparticles in the flat band as expected from the fine-tuning construction of the model Hamiltonian (\ref{normalized_Hamiltonian}).
Here, in the limit $\eta\gg |\Delta_{1,2}|$, the lower band dispersion is given by
\begin{equation}
E_{{\bm k},-} = \pm \frac{|\Delta_1 + \Delta_2 \lambda^2 k^2|}{1+\lambda^2 k^2}.
\end{equation}
Due to momentum-dependent interband hybridization, the term $\Delta_{1}$ opens up a gap in the spectrum of flat-band quasiparticles at small momenta $\lambda k \ll1$, where contribution of the upper band is suppressed, while the term $\Delta_{2}$ in the spectral gap becomes significant at large momenta, $\lambda k \gg 1$.

We emphasize that when $\Delta_1$ and $\Delta_2$ have a $\pi$-phase shift, the spectrum remains gapped. It can be seen from (\ref{Dispersion_SC}), after taking into account the terms $ \propto O(|\Delta_{1,2}|^2/\eta)$ in the expansion at $\lambda k = |\Delta_1/\Delta_2|$ in this particular case. The spectral gap reaches its maximum when the phase difference between $\Delta_1$ and $\Delta_2$ vanishes. We assume this condition in what follows.

\subsection{Phase stiffness}

One can further examine the superconducting phase stiffness in the system through the calculation of the superconducting current density. The current operator in the static approximation is ${\bm J}({\bm r}) = 2 e \eta \lambda^2 [-i \boldsymbol{\partial}_r- (e/c) {\bm A}({\bm r})\tau_3] (1+\sigma_3)/2 + e \eta \lambda  \boldsymbol{\sigma}$, where $e<0$ is the electron charge, $c$ is the speed of light, and ${\bm A}$ is the vector potential. Pauli matrix $\tau_3$  act in the Nambu space.

%%%%%%%%%%%%%%%%%%
%%%%%%%%%%%%%%%%

Following standard textbook derivations \cite{AGD}, utilizing the Matsubara Green function in Nambu representation which can be found inverting the matrix $G({\bm k},\omega_n) = [i\omega_n - H_{\rm BdG}({\bm k})]^{-1}$, in the London limit the superconducting current density, linear in the vector potential, can be written as
\begin{equation}
{\bm J}({\bm r}) = - \frac{4 e^2}{\hbar^2 c} D {\bm A}({\bm r}),
\end{equation}
where $D$ is the diagonal component of superconducting phase stiffness matrix and $\hbar$ is explicitly restored here for dimensional clarity. We note that transverse supercurrent response vanishes in this static and long-wave limit. However, if the time reversal symmetry is broken, off-diagonal components of the superconducting stiffness matrix become finite giving rise to the Edelstein effect in flat-band superconductor, \cite{Edelstein}.

In the perfectly symmetric scenario $|\Delta_1| = |\Delta_2|$, the flat band is completely gapped. As a result, transport properties at low temperatures are gap-dominated and the temperature dependence of the superfluid stiffness is exponentially suppressed at $T\ll |\Delta_2|$, similarly as in the case of superconductivity in graphene and
of surface superconductivity of a flatband multilayered rhombohedral graphene \cite{Kopnin_Sonin_correction, Kopnin_MAIN}.

The effect of momentum dependence of the band dispersion on the phase stiffness is more pronounced in the limit when one of the gaps is much larger than the other.
In the asymmetric case it is convenient to consider the limit where $\Delta_1=0$ and $\Delta_2\neq 0$ (in the opposite case when $\Delta_2=0$ and $\Delta_1 \neq 0$, we obtain same results for the phase stiffness applying a formal substitution $|\Delta_2| \rightarrow |\Delta_1|$). The band dispersion exhibits a parabolic node, scaling as $\propto k^2$ at small momenta, and saturates to a momentum-independent value $|\Delta_2|$ for $\lambda k \gg1$. After subtracting the normal-state contribution, analogously to the procedure used for superconductivity in graphene \cite{Kopnin_Sonin_correction, Kopnin_MAIN}, the phase stiffness is
\begin{equation}
D = \frac{T^2}{\pi |\Delta_2|} \int_0^{\frac{|\Delta_2|}{2T}} x {\rm th}(x) dx.
\end{equation}
Formally, taking the low gap limit, $|\Delta_2|/T < 1$, yields $D= |\Delta_2|^2/(24 \pi T)$. At low temperatures $T/|\Delta_2| \ll 1$, we find
\begin{equation}
D = \frac{|\Delta_2|}{8 \pi}\left( 1- \frac{\pi^2 T^2}{3|\Delta_2|^2}\right).
\end{equation}
The phase stiffness is linearly proportional to the gap amplitude, similarly to the result obtained for flat-bands in multilayer graphene and graphene at charge neutrality \cite{Kopnin_Sonin_correction, Kopnin_MAIN}.
The gap node results in the quadratic temperature dependence of the superfluid stiffness at charge neutrality and at low temperatures.  

We note that the parabolic node vanishes as the doping increases.
For example, at $\eta\gg |\mu|>|\Delta_2|$, we obtain
\begin{equation}
E_{\bm k, -} = \pm \sqrt{\mu^2 + \frac{[\lambda^2 k^2- (\mu/\eta)]^2 - (\mu/\eta)^2}{(1+ \lambda^2 k^2)^2} |\Delta_2|^2}.
\end{equation}
The doping suppresses the amplitude as well as the temperature dependence of the phase stiffness, which is given by $D = (|\Delta_2|^2/12 \pi |\mu|)(1- 2e^{-|\mu|/T})$ at low temperatures. 

To obtain a comparative scale for phase ordering, we adopt the formal BKT criterion as a working approximation, $T_{\rm BKT} = (\pi/2) D|_{T=T_{\rm BKT}}$, yielding $T_{\rm BKT}/|\Delta_2| =1/16$. Within this model, the resulting temperature dependence of the amplitude of $D$ is therefore negligible. We believe it cannot be applied to explain temperature dependence of the superfluid stiffness observed in experiments for graphene-based structures \cite{Tanaka_2025, Banerjee_2025}. It should be also noted that $T_{\mathrm{BKT}}$ is to be regarded only as an upper bound for possible phase ordering, without implying the actual realization of a BKT transition in flat-band.

\section{Discussion and Conclusion}
Now let us comment on the effect of nonmagnetic and intraband disorder on the stability of the Cooper pairing in such flat band system.
The theory of an Anderson impurity effect on the subgap resonances in the case of Cooper pairing in flat band limit, where the BCS Hamiltonian is taken in the zero bandwidth limit, so that the superconducting system may be considered as a single energy level with onsite pairing, was developed in Ref. \cite{Zitko_PhysRevB}. To gain qualitative insight in our two-band case, we consider an Anderson model of single short-range, non-magnetic impurity neglecting on-site Coulomb repulsion.

In conventional s-wave superconductors, such impurity produces single-quasiparticle subgap resonances near the gap edges and often negligible \cite{Machida_Shibata, Balatsky_RevModPhys}. However, the flat-band system may provide conditions where the pairing gap amplitude is larger than the chemical potential for these resonances to lie deep inside the gap.

The localized state with energy $E_{\rm d}$ (measured relative to the Fermi level of the flat-band electrons) hybridizes with a superconducting flat-band via potential $V_{\rm sd}>0$. 
The poles of the t-matrix,
\begin{equation}
t(\omega_n) = V_{\rm sd}^2\tau_{3}\left[i\omega_n - E_{\rm d} \tau_3 - V_{\rm sd}^2\int_{\bm k} \tau_3 G({\bm k},\omega_n)\tau_3\right]^{-1}\tau_3,
\end{equation}
after taking the analytical continuation, $t(\omega_n)|_{i \omega_n\rightarrow \omega+i0+}$, determine the energies of these bound states. Focusing on the flat-band contribution, we obtain
\begin{eqnarray}\nonumber\label{T_matrix}
2\omega^2 &=& |\Delta_2|^2+E_{\rm d}^2+2\Gamma^2\\
&\pm& \sqrt{(|\Delta_2|^2-E_{\rm d}^2)^2 + 4 \Gamma^2(|\Delta_2|^2+E_{\rm d}^2)},
\end{eqnarray}
where $\Gamma = V_{\rm sd} k_0/\sqrt{4\pi}$ parametrizes the hybridization strength in terms of the flat band density of states determined by $k_0$.  This result is generic for the superconductivity in dispersionless flat-band system in agreement with Ref. \cite{Zitko_PhysRevB}. 

To analyze (\ref{T_matrix}), consider the strong-hybridization limit, $\Gamma > |\Delta_2|, |E_{\rm d}|$. Large $\Gamma$ tends to push resonances out of the gap, with $|\omega| \propto \Gamma$. This behavior differs from that in conventional s-wave superconductors, where strong hybridization results in bound states lying at the gap edge, \cite{Machida_Shibata, Balatsky_RevModPhys}.
In the weak hybridization case, $\Gamma \ll |\Delta_2|, |E_{\rm d}|$, if the impurity level lies outside the gap $|E_{\rm d}|>|\Delta_2|$, one obtains $\omega^2 = |\Delta_2| - 2\Gamma^2|\Delta_2|^2 /(E_{\rm d}^2- |\Delta_2|^2)$. The resonance approaches the gap edge from below, describing a shallow subgap state. However, if the impurity level lies inside the gap, $|E_{\rm d}|<|\Delta_2|$, the subgap resonance is located at $\omega^2 = E_{\rm d}^2 - 2\Gamma^2E_{\rm d}^2/(|\Delta_2|^2 - E_{\rm d}^2) + \Gamma^4(|\Delta_2|^2 + E_{\rm d}^2)^2/(|\Delta_2|^2 - E_{\rm d}^2)^3$. This solution continuously connects to the bare impurity energy $E_{\rm d}$ as $\Gamma\rightarrow 0$ and remains a subgap resonance. In the special case $E_{\rm d}=0$, the subgap energy scales as $
|\omega| = \Gamma^2/|\Delta_2| \propto |\Delta_2| \left(V_{\rm sd}/V_2\right)^2$. The subgap resonances reflect sensitivity of flat-band superconductivity to disorder as compared with the conventional superconductor.  

To conclude, the main result of this work is the identification of a nodal superconducting regime in two-band system arising when a dispersive band hybridizes with a nearly flat one with momentum-dependent interband mixing. We adopted the model of preformed Cooper pairs, where the large density of states in the flat band enhances pair formation which may lead to a separation between the pairing and phase coherence temperature scales. A characteristic feature of the model is a parabolic node in the quasiparticle spectrum when pairing is band-asymmetric. This nodal structure produces a quadratic temperature dependence of the phase stiffness.

\acknowledgements
We thank E. V. Blinova and M. Kumar for support and acknowledge the Pirinem School of Theoretical Physics. 

%%%%%%%%%%%%%%%%%%%%%%%%%%%%%%%%%%%%%%%%%%%%%%%%%%%%%%%%%%%%%%%%%%
%%%% Bibliography
%\bibliographystyle{apsrev}
\bibliography{TopoFBSC_references_2}
%%%%%%%%%%%%%%%%%%%%%%%%%%%%%%%%%%%%%%%%%%%%%%%%%%%%%%%%%%%%%%%%%%

\end{document}